\documentclass{elsart}
\usepackage{amssymb}

\renewcommand{\theequation}{\arabic{section}.\arabic{equation}}

\newcommand{\vek}[1]{\mbox{\boldmath$#1$}}

\newcommand{\nablaleftright}{\stackrel{\leftrightarrow}{\nabla}}
\newcommand{\nablaleft}{\stackrel{\leftarrow}{\nabla}}
\newcommand{\nablaright}{\stackrel{\rightarrow}{\nabla}}

\def\trans{\mbox{\tiny$\bot$}} 
\def\longi{\mbox{\tiny$\|$}}   

\begin{document}

\begin{frontmatter}
\title{Kinetic theory of QED plasmas in a strong electromagnetic field\\
II. The mean-field description}

\author[Rostock]{A. H\"oll\thanksref{Arne}}
\author[Moscow]{V.G. Morozov\thanksref{Vladimir}}
\author[Rostock]{G. R\"opke\thanksref{Gerd}}
\address[Rostock]{Physics Department, University of Rostock,
Universit\"atsplatz 3, D-18051 Rostock, Germany}
\address[Moscow]{Moscow State Institute of Radioengineering,
Electronics, and Automation, 117454 Vernadsky Prospect 78, Moscow, Russia}
\thanks[Arne]{hoell@darss.mpg.uni-rostock.de}
\thanks[Vladimir]{vmorozov@orc.ru}
\thanks[Gerd]{gerd@darss.mpg.uni-rostock.de}

\begin{abstract}
Starting from a general relativistic kinetic equation, a self-consistent
mean-field equation for fermions is derived within a
covariant density matrix approach of QED plasmas in strong external
fields. A Schr\"odinger picture formulation on space-like
hyperplanes is applied.
The evolution of the distribution function is described by the
one-particle gauge-invariant $4\times 4$ Wigner matrix, which is decomposed in spinor space.
A coupled system of equations for the corresponding Wigner 
components is obtained. The polarization current is expressed in terms of the
Wigner function. Charge conservation is obeyed.
In the quasi-classical limit for the Wigner components a relativistic
Vlasov equation is obtained, which is presented in an invariant,
i.e. hyperplane independent, form.
\end{abstract}

\end{frontmatter}

\setcounter{equation}{0}
\section{Introduction}

In part one of this article~\cite{Hoell1_01} we have developed
a covariant density matrix approach to kinetic theory of
QED plasmas, making use of the
relativistic hyperplane formalism~\cite{Bogoliubov51,Fleming65,Fleming66}
in the Schr\"odinger picture.
In what follows the paper~\cite{Hoell1_01} will be
referred to as I and equations from this paper will be labeled by (I$\ldots$),
where ``$\ldots$'' denotes the equation number.
In the present paper we aim to derive quantum mean-field kinetic equations for the
fermionic subsystem starting from a general relativistic covariant equation
discussed in paper I.

Section~2 briefly recalls notations and definitions from paper I needed
in further considerations. Section~3 is devoted to the derivation of a
mean-field kinetic equation for the gauge-invariant
fermionic Wigner function in the
hyperplane formalism. We next use the spinor decomposition of the Wigner
matrix and obtain a set of coupled covariant equations describing kinetic
processes in
different channels. The charge conservation is shown to be fulfilled.
In Section~4 we discuss the quasi-classical limit in the hyperplane formalism.
We derive mean-field kinetic equations for the distribution functions
of particles and antiparticles, and show that these equations can be
represented in a fully covariant form. Section~5 will conclude and
give a short outlook.
In Appendix~A we show the relation between our approach and the existing
mean-field theories of QED~\cite{BGR91,ShinRafelski}.
Finally, Appendix~B gives expressions for the
matrix commutation and anticommutation relations which are necessary for the
spinor decomposition of the kinetic equation.

Except for the quasi-classical limit, we use the system of units with
$\hbar=c=1$. The signature of the metric tensor is $(+,-,-,-)$.

\setcounter{equation}{0}
\section{Basic definitions}

We consider a quantum plasma of charged fermions interacting through the
electromagnetic (EM) field. For simplicity, we will
take these fermions to be electrons and positrons, but the inclusion other
fermions (say, protons) as additional Dirac fields is not a particular
problem. The system is assumed to be subjected to a prescribed external EM
field which is not necessarily weak.

In paper I we defined the one-particle density matrix for fermions as the
average
\begin{equation}
\label{II:DensMatr:Def} \rho^{}_{aa'}
\left(x^{}_{\trans}, x^{\prime}_{\trans};\tau\right)
= \left\langle \hat{\rho}^{}_{aa'} \left(x^{}_{\trans},
x^{\prime}_{\trans}\right) \right\rangle^{\tau}\equiv {\rm Tr}\left\{
\hat{\rho}^{}_{aa'}
\left(x^{}_{\trans}, x^{\prime}_{\trans}\right)\, \varrho(n,\tau)
\right\},
\end{equation}
where $\varrho(n,\tau)$ is the nonequilibrium statistical operator of the
system on a hyperplane $\sigma^{}_{n,\tau}$  characterized by a unit
time-like normal four-vector $n^{\mu}$ and a scalar parameter $\tau=x\cdot n$
which may be interpreted as an ``invariant time''. The fermionic density
operator $\hat{\rho}$ is given in the Schr\"odinger picture on the
hyperplane by
\begin{equation}
\label{II:DensOper:Com}
\hat{\rho}^{}_{aa'}(x^{}_{\trans},x^{\prime}_{\trans})=
-{1\over2}\big[ \hat\psi^{}_{a}(x^{}_{\trans}), \,\hat{\!\bar
\psi}^{}_{a'}(x^{\prime}_{\trans}) \big],
\end{equation}
$a,a'$ being spinor indices of the Dirac field operators. The transverse
components (which are space-like) of $x=\{x^{\mu}\}$ and other four-vectors
$V=\{V^{\mu}\}$ are defined with respect to the  normal $n$ through the
decomposition
\begin{equation}
\label{II:DecompXV} x^{\mu}_{}=n^{\mu}\tau + x^{\mu}_{\trans}, \qquad
 V^{\mu}=n^{\mu}V^{}_{\longi} + V^{\mu}_{\trans},
\end{equation}
where
\begin{equation}
\label{II:Trans} V^{}_{\longi}=n\cdot V, \qquad
V^{\mu}_{\trans}=\Delta^{\mu}_{\ \nu} V^{\nu}, \qquad \Delta^{\mu}_{\ \nu}=
\delta^{\mu}_{\ \nu} - n^{\mu} n^{}_{\nu}.
\end{equation}
Our further analysis rests heavily on the basic ``equal-time''
anticommutation relations for the Dirac field operators on hyperplanes
[cf.~Eqs.(I.3.24) and~(I.3.25)]
\begin{equation}
\label{II:Anticomm}
\begin{array}{l}
\displaystyle \bigg\{ \hat{\psi}_{a}^{} (\tau ,x_{\trans}^{}),
\,\hat{\!\bar\psi}_{\!a'} (\tau ,x_{\trans}^{\prime}) \bigg\}
=\left[\gamma^{}_{\longi}(n)\right]^{}_{aa'} \delta_{}^{3}(x_{\trans}^{} -
x_{\trans}^{\prime}),
\\[10pt]
\displaystyle \bigg\{ \hat{\psi}_{a}^{} (\tau ,x_{\trans}^{}),
\,\hat{\psi}^{}_{a'} (\tau ,x_{\trans}^{\prime}) \bigg\} = \bigg\{
\hat{\!\bar\psi}_{\!a}^{} (\tau ,x_{\trans}^{}), \,\hat{\!\bar\psi}_{\!a'}
(\tau ,x_{\trans}^{\prime}) \bigg\} =0,
\end{array}
\end{equation}
where
\begin{equation}
\label{II:DeltaFunc} \delta^{3}(x^{}_{\trans})= \int \frac{d^4 p}{(2\pi)^3}\,
{\rm e}^{-ip\cdot x}\, \delta(p\cdot n)
\end{equation}
is the three-dimensional delta function on a hyperplane $\sigma^{}_{n,\tau}$ and
the matrix $\gamma^{}_{\longi}(n)$ is defined through the following
decomposition of the Dirac matrices:
\begin{equation}
\label{II:GammaDecomp}
\begin{array}{c}
\gamma^{\mu}=n^{\mu}\gamma^{}_{\longi}(n) + \gamma^{\mu}_{\trans}(n),
\\[6pt]
\gamma^{}_{\longi}(n)=n^{}_{\mu} \gamma^{\mu}, \qquad
\gamma^{\mu}_{\trans}(n)= \Delta^{\mu}_{\ \nu} \gamma^{\nu}.
\end{array}
\end{equation}

As discussed in paper I, the self-consistent mean-field approximation for the
fermionic subsystem can be introduced only when the EM field variables are
separated into the macroscopic condensate mode and the photon degrees of
freedom. This we have shown in paper I by means of a time-dependent unitary
transformation of the statistical operator and the operators of the EM
field [see Eqs.~(I.4.1) and~(I.4.3)].
After this procedure, the effective Hamiltonian describing the fermionic
subsystem in the mean-field approximation can be taken in the form
\begin{equation}
\label{II:EffHam} \hat{\mathcal H}^{\tau}_{0}(n) = \int_{\sigma^{}_{n}}
d\sigma\, \,:\hat{\!\bar \psi} \left( -\frac{i}{2} \gamma^{\mu}_{\trans}(n)
\nablaleftright^{}_{\mu} +
 m \right)\hat{\psi}:
+\int\limits_{\sigma^{}_{n}} d\sigma\,
\,\hat{\!j}^{}_{\mu}(x^{}_{\trans})\,{\mathcal A}^{\mu}(\tau,x^{}_{\trans}),
\end{equation}
where the symbol $:\!\hat{O}\!:$ shows the normal ordering in
operators, and the space-like derivatives
$\nablaleftright^{}_{\mu}=\nablaright^{}_{\mu}-\nablaleft^{}_{\mu}$
are defined by the relations
\begin{equation}
\label{II:Derivat}
\partial^{}_{\mu}= n^{}_{\mu}\,\frac{\partial}{\partial\tau}
+ \nabla^{}_{\mu}, \qquad \nabla^{}_{\mu}= \Delta_{\mu}^{\ \nu}
\partial^{}_{\nu} = \Delta_{\mu}^{\ \nu}\,\frac{\partial}{\partial
x^{\nu}_{\trans}}.
\end{equation}
We will use the notation
$G(\tau,x^{}_{\trans})\equiv G(n\tau + x^{}_{\trans})$ for any
function $G(x)$ on the hyperplane $\sigma^{}_{n,\tau}$ furthermore.

The first term in Eq.~(\ref{II:EffHam}) is the Hamiltonian of free fermions,
while the second term describes their interaction with the total mean EM
field in the system, ${\mathcal A}^{\mu}$. We have shown in paper I that the
total field tensor ${\mathcal F}^{\mu\nu}=\partial^{\mu}{\mathcal A}^{\nu} -
\partial^{\nu}{\mathcal A}^{\mu}$ satisfies Maxwell equations
\begin{equation}
\label{II:Maxwell:Tot}
\partial^{}_{\mu} {\mathcal F}^{\mu\nu}(x)=j^{\nu}(x) +
j^{\nu}_{\rm ext}(x).
\end{equation}
Here $j^{\nu}_{\rm ext}(x)$ is a prescribed external current and
\begin{equation}
\label{II:PolCurr} j^{\mu}(x)=\langle
\,\,\hat{\!j}^{\,\mu}(x^{}_{\trans})\rangle^{\tau}
\end{equation}
is the mean polarization current. For the electron-positron plasma, the
current operator is ($e<0$)
\begin{equation}
\label{II:OpCurr}
  \hat{\!j}^{\,\mu}(x^{}_{\trans})= e
:\hat{\!\bar\psi}(x^{}_{\trans}) \gamma^{\mu} \hat{\psi}(x^{}_{\trans})\!: .
\end{equation}
If protons are treated as a dynamical subsystem, the corresponding term must
be included into the current operator.

As outlined in paper I, the total Hamiltonian of the system contains, in addition
to~(\ref{II:EffHam}), the Hamiltonian of free photons, $\hat{H}^{}_{EM}(n)$,
and the term $\hat{\mathcal H}^{\tau}_{\rm int}(n)$,
describing the interaction between fermions and photons.
In the mean-field approximation the interaction term is neglected.
The derivation of the fermionic kinetic equation requires to calculate 
commutators of the dynamical fermion operators $\hat{\psi}$ and 
$\hat{\!\bar\psi}$ with the Hamiltonian (see below). The free photon
contribution  $\hat{H}^{}_{EM}(n)$, consisting completely of the
photon dynamical operators $\partial^\nu \hat{A}_{\trans}^{\mu}$, will
not contribute to these commutators and can therefore be omitted.
This implies that the dynamics of the EM field in the mean-field
approximation is completely governed by the Maxwell
equations~(\ref{II:Maxwell:Tot}). Nevertheless, at the end of the
paper we shall discuss some non-trivial relations between the mean-field
description of the fermionic subsystem and the photon kinetics in QED
plasmas.

To complete the list of definitions, we write down the expression for the
gauge-invariant ``one-time'' Wigner function~\cite{Elze1_87} on the hyperplane
$\sigma^{}_{n,\tau}$. The Wigner function is expressed in terms of the
one-particle density matrix~(\ref{II:DensMatr:Def}) by
\begin{eqnarray}
\label{II:Wigner:Def} & & \hspace*{-20pt}
W^{}_{aa'}(x^{}_{\trans},p^{}_{\trans};\tau)= \int d^4 y\, {\rm e}^{ip\cdot
y}\,\delta(y\cdot n)\,
\nonumber\\[6pt]
& &
{}\times \exp\left\{ ie\Lambda(x^{}_{\trans} +\mbox{$1\over2 $}y^{}_{\trans},
x^{}_{\trans}-\mbox{$1\over2 $}y^{}_{\trans};\tau) \right\}
\rho^{}_{aa'}
\left(x^{}_{\trans}+\mbox{$1\over2 $}y^{}_{\trans},
x^{}_{\trans}-\mbox{$1\over2 $}y^{}_{\trans};\tau\right),
\end{eqnarray}
where
\begin{eqnarray}
\label{II:GaugeFunc} \Lambda(x^{}_{\trans},x^{\prime}_{\trans};\tau) &=&
\int\limits_{x^{\prime}_{\trans}}^{x^{}_{\trans}} {\mathcal
A}^{}_{\trans\mu}(\tau,R^{}_{\trans})\, dR^{\mu}_{\trans}
\nonumber\\[6pt]
{}&\equiv& \int\limits_{0}^{1} ds \left(x^{\mu}_{\trans} -
x^{\prime\mu}_{\trans}\right) {\mathcal A}^{}_{\trans\mu}\big( \tau,
x^{\prime}_{\trans} +s(x^{}_{\trans} - x^{\prime}_{\trans}) \big)
\end{eqnarray}
is the gauge function. Our immediate task will be to derive a mean-field
kinetic equation for $W$.

\setcounter{equation}{0}

\section{Mean-field kinetic equations}

\subsection{Kinetic equation for the one-particle density matrix}
We start with the mean-field kinetic equation for the density
matrix~(\ref{II:DensMatr:Def}). Taking~(\ref{II:EffHam}) as the effective
Hamiltonian for the fermionic subsystem, we have
\begin{equation}
 \label{II:FDens:EqMot}
\frac{\partial}{\partial\tau}\,
\rho^{}_{aa'}(x^{}_{\trans},x^{\prime}_{\trans};\tau)
= -i\, \left\langle \big[
\hat{\rho}^{}_{aa'}
(x^{}_{\trans},x^{\prime}_{\trans}), \hat{\mathcal H}^{\tau}_{0}(n)\big]
\right\rangle^{\tau}.
\end{equation}
The commutator in the right-hand side is easily calculated by using the
identity
\begin{eqnarray}
\label{II:CommF&N} & & \hspace*{-20pt} \big[
\hat{\rho}^{}_{aa'}(x^{}_{\trans},x^{\prime}_{\trans}),
:\hat{\!\bar\psi}^{}_{b}(y^{}_{\trans})
\hat{\psi}^{}_{b'}(y^{\prime}_{\trans})\!:\big]= (\gamma^{}_{\longi})^{}_{ab}
\, \delta^{3}(x^{}_{\trans}-y^{}_{\trans}) \,
\hat{\rho}^{}_{b'a'}(y^{\prime}_{\trans},x^{\prime}_{\trans})
\nonumber\\[6pt]
& & \hspace*{140pt} {}- (\gamma^{}_{\longi})^{}_{b'a'}\,
\delta^{3}(x^{\prime}_{\trans}-y^{\prime}_{\trans}) \,
\hat{\rho}^{}_{ab}(x^{}_{\trans},y^{}_{\trans}) ,
\end{eqnarray}
which follows from the anticommutation relations~(\ref{II:Anticomm}). After
some algebra we find that Eq.~(\ref{II:FDens:EqMot}) can be written in matrix
notation
$\rho \equiv \left[ \rho^{}_{aa'} \right]$ as
\begin{eqnarray}
\label{II:DensM:KEq} \frac{\partial}{\partial\tau}
\rho(x^{}_{\trans},x^{\prime}_{\trans};\tau)
&=& -im\left[\gamma^{}_{\longi},
\rho (x^{}_{\trans},x^{\prime}_{\trans};\tau) \right]
\nonumber\\[6pt]
& & {}+\left( -i\nabla^{}_{\mu} +e{\mathcal
A}^{}_{\trans\mu}(\tau,x^{}_{\trans}) \right) S^{\mu}
\rho(x^{}_{\trans},x^{\prime}_{\trans};\tau)
\nonumber\\[6pt]
& & {}+ \left( i\nabla^{\prime}_{\mu} + e{\mathcal
A}^{}_{\trans\mu}(\tau,x^{\prime}_{\trans})\right)
\rho(x^{}_{\trans},x^{\prime}_{\trans};\tau) S^{\mu}
\nonumber\\[6pt]
& & {}-ie\left( {\mathcal A}^{}_{\longi}(\tau,x^{}_{\trans}) - {\mathcal
A}^{}_{\longi}(\tau,x^{\prime}_{\trans})
\right)\rho(x^{}_{\trans},x^{\prime}_{\trans};\tau) ,
\end{eqnarray}
where
\begin{equation}
\label{II:SMatr} S^{\mu}= \bar{\sigma}^{\mu\nu} n^{}_{\nu}, \qquad \bar{\sigma}^{\mu\nu}=
{i\over2} \left[\gamma^{\mu},\gamma^{\nu}\right].
\end{equation}
Eq.~(\ref{II:DensM:KEq}) defines the mean-field dynamics of the fermionic
one-particle density matrix for time-like translations with respect to the
space-like plane $\sigma_{n,\tau}^{}$. In the special Lorentz frame where
$n^{\mu}=(1,0,0,0)$ (which sometimes will be referred to as the ``instant
frame''), we have $\tau=t$ and therefore Eq.~(\ref{II:DensM:KEq}) describes
the time evolution of the one-particle density matrix. It is interesting to
note that, within the mean-field description, there is no need to know the
explicit form of the nonequilibrium statistical operator $\varrho(n,\tau)$.
The mean-field kinetic equation follows directly from the equation of motion
for the density operator $\hat{\rho}$ with the effective
Hamiltonian~(\ref{II:EffHam}).

\subsection{Kinetic equation for the Wigner function}
Applying the Wigner transformation~(\ref{II:Wigner:Def}) to
Eq.~(\ref{II:DensM:KEq}), we obtain in matrix notation $W \equiv \left[
W_{aa'}^{}\right]$
\begin{equation}
\label{II:Wigner:Eq} {\sf D}^{}_{\tau}W=- im\big[\gamma^{}_{\longi},W\big]
-{i\over2}\,{\sf D}^{}_{\trans\mu} \left[S^{\mu},W\right] - {\sf P}^{}_{\mu}
\left\{S^{\mu},W\right\},
\end{equation}
where we have introduced the operators
\begin{eqnarray}
& & \label{II:DTau} {\sf D}^{}_{\tau}= \frac{\partial}{\partial\tau}
-e\int\limits^{1/2}_{-1/2} ds\, n^{\mu} {\mathcal F}^{}_{\mu\nu}
\left(\tau,x^{}_{\trans} -is\nabla^{}_{p}\right) \nabla^{\nu}_{p},
\\[6pt]
& & \label{II:DTrans} {\sf D}^{}_{\trans\mu}= \nabla^{}_{\mu}
-e\int\limits_{-1/2}^{1/2} ds\, {\mathcal F}^{}_{\trans\mu\nu}
\left(\tau,x^{}_{\trans} -is\nabla^{}_{p}\right) \nabla^{\nu}_{p},
\\[6pt]
& & \label{II:PTrans} {\sf P}^{}_{\mu}= p^{}_{\trans\mu} -ie
\int\limits_{-1/2}^{1/2} s\,ds\, {\mathcal F}^{}_{\trans\mu\nu}
\left(\tau,x^{}_{\trans} -is\nabla^{}_{p}\right) \nabla^{\nu}_{p},
\end{eqnarray}
and the transverse gradient in the momentum space:
$\nabla^{\mu}_{p}=\Delta^{\mu\nu} \,\partial/
\partial{p^{\nu}_{\trans}}$.
The transverse part of the total field tensor is defined as
\begin{equation}
\label{II:TransFTens} {\mathcal F}^{\mu\nu}_{\trans}= \nabla^{\mu} {\mathcal
A}^{\nu}_{\trans} - \nabla^{\nu} {\mathcal A}^{\mu}_{\trans}.
\end{equation}
The virtue of Eq.~(\ref{II:Wigner:Eq}) is its compact and covariant form. It
should be emphasized, however, that the Wigner function, the matrices
$\gamma^{}_{\longi}$, $S^{\mu}$, and the
operators~(\ref{II:DTau})~--~(\ref{II:PTrans}) are defined with respect to
the family of hyperplanes $\sigma^{}_{n,\tau}$ characterized by the normal
$n^{\mu}$. The fact that $n^{\mu}$ is an arbitrary time-like unit vector
reflects Lorentz covariance of Eq.~(\ref{II:Wigner:Eq}). To see this, we note
that the normal vectors of the same plane in different frames,
$n^{\mu}$ and $n^{\prime \mu}$, are related by a Lorentz transformation
(boost) $n^{\prime \mu}=\Lambda^{\mu}_{\ \nu} n^{\nu}$ which is equivalent to
the transformation of space-time coordinates $x^{\prime\mu}=\Lambda^{\mu}_{\
\nu} x^{\nu}$. In the new Lorentz frame the invariant time parameter $\tau$
has the same value, since $\tau=n\cdot x=n^{\prime}\cdot x^{\prime}$. For any
given $n$, there exists the special ``instant frame'' where
$n^{\mu}=(1,0,0,0)$ and, consequently, $\tau=t$. In Appendix~A we show that
in this frame Eq.~(\ref{II:Wigner:Eq}) reduces to the kinetic equation
derived by Bialynicki-Birula et al.~\cite{BGR91} on the basis of a different
approach.

Despite its apparently simple structure, Eq.~(\ref{II:Wigner:Eq}) is a rather
complicated matrix equation. To obtain a deeper physical insight into
processes described by this equation, it is convenient to expand the Wigner
function in a complete basis in spinor space
\begin{equation}
\label{II:Wigner:Decomp} W={1\over4}\left( I\mathcal{W} + \gamma^{}_{\mu}
\mathcal{W}^{\mu}_{} + \gamma^{}_{5} \mathcal{W}^{}_{(P)} +
\gamma^{}_{5}\gamma^{}_{\mu} \mathcal{W}^{\mu}_{(A)} + \bar{\sigma}^{}_{\mu\nu}
\mathcal{W}^{\mu\nu}_{} \right) .
\end{equation}
Here $I$ is the unit matrix, and $\mathcal{W}^{\mu}_{}$,
$\mathcal{W}^{}_{(P)}$, $\mathcal{W}^{\mu}_{(A)}$, $\mathcal{W}^{\mu\nu}_{}$
are the scalar, vector, pseudo-scalar, axial-vector and tensor coefficient
function of the Wigner matrix $W$ respectively. By the trace rules in spinor
space it can easily be verified that the coefficient functions can be
expressed as
\begin{eqnarray}
\label{II:Wig:Scalar} & & \mathcal{W}_{}^{} = \rm{tr}\left( W \right) ,
\\[4pt]
\label{II:Wig:Vector} & & \mathcal{W}_{}^{\mu} = \rm{tr}\left(
\gamma_{}^{\mu}W \right) ,
\\[4pt]
\label{II:Wig:PScalar} & & \mathcal{W}_{(P)}^{} = \rm{tr}\left(
\gamma_{5}^{}W \right) ,
\\[4pt]
\label{II:Wig:AVector} & & \mathcal{W}_{(A)}^{\mu} = \rm{tr}\left(
\gamma_{}^{\mu}\gamma_{5}^{}W \right) ,
\\
\label{II:Wig:Tensor} & & \mathcal{W}_{}^{\mu\nu} = \frac{1}{2} \rm{tr}
\left( \bar{\sigma}_{}^{\mu\nu}W \right),
\end{eqnarray}
where the symbol ``tr'' means the trace over spinor indices. Now a
straightforward algebra allows to derive a coupled set of equations for the
coefficient functions from the kinetic equation~(\ref{II:Wigner:Eq}) by using
Eqs.~(\ref{II:Wig:Scalar})~--~(\ref{II:Wig:Tensor})
\begin{eqnarray}
\label{II:Eq:Scalar} & & \hspace*{-25pt} {\sf D}^{}_{\tau}
\mathcal{W}^{}_{}=2 \Big( n^{}_{\alpha}{\sf P}^{}_{\beta} - n^{}_{\beta} {\sf
P}^{}_{\alpha} \Big) \mathcal{W}^{\alpha\beta}_{},
\\[4pt]
\label{II:Eq:Vector} & & \hspace*{-25pt} {\sf D}^{}_{\tau}
\mathcal{W}^{\mu}_{}= - \Big( n^{\mu} {\sf D}^{}_{\!\trans\alpha}-
n^{}_{\alpha}{\sf D}^{\mu}_{\!\trans} \Big) \mathcal{W}^{\alpha}_{} -
2\varepsilon^{\mu}_{\ \alpha\beta\lambda}\, n^{\alpha}\, {\sf P}^{\beta}
\mathcal{W}^{\lambda}_{(A)^{}} - 4m
 \mathcal{W}^{\mu\alpha} n^{}_{\alpha} ,
\\[4pt]
\label{II:Eq:PScalar} & & \hspace*{-25pt} {\sf D}^{}_{\tau}
\mathcal{W}^{}_{(P)}= 2im n^{}_{\alpha} \mathcal{W}^{\alpha}_{(A)^{}} +
2i\varepsilon^{}_{\alpha\beta\lambda\varrho}\, n^{\alpha} {\sf P}^{\beta}
\mathcal{W}^{\lambda\varrho},
\\[4pt]
\label{II:Eq:AVector} & & \hspace*{-25pt} {\sf D}^{}_{\tau}
\mathcal{W}^{\mu}_{(A)^{}}= - 2\varepsilon^{\mu}_{\ \alpha\beta\lambda}\,
n^{\alpha} {\sf P}^{\beta} \mathcal{W}^{\lambda} + 2im n^{\mu}
\mathcal{W}^{}_{(P)} - \Big(n^{\mu} {\sf D}^{}_{\!\trans\alpha} -
n^{}_{\alpha} {\sf D}^{\mu}_{\!\trans} \Big) \mathcal{W}^{\alpha}_{(A)^{}},
\\[4pt]
\label{II:Eq:Tensor} & & \hspace*{-25pt} {\sf D}^{}_{\tau}
\mathcal{W}^{\mu\nu}_{}= \Big( n_{}^{\mu}{\sf P}^{\nu}_{} - n_{}^{\nu}{\sf
P}^{\mu}_{} \Big) \mathcal{W}^{}_{} - m \Big( n^{\mu}_{} \mathcal{W}^{\nu}_{}
- n^{\nu}_{} \mathcal{W}^{\mu}_{} \Big) +
i\varepsilon^{\mu\nu\alpha\beta}_{}\, n^{}_{\alpha} {\sf P}^{}_{\beta}
\mathcal{W}^{}_{(P)}
\nonumber\\[2pt]
& & \hspace*{-25pt} \hspace*{60pt} {}+ \,\Big( n^{\mu}_{} {\sf
D}^{}_{\!\trans\alpha} - n^{}_{\alpha} {\sf D}^{\mu}_{\!\trans} \Big)
\mathcal{W}^{\nu\alpha}_{} - \Big( n^{\nu}_{} {\sf D}^{}_{\!\trans\alpha} -
n^{}_{\alpha} {\sf D}^{\nu}_{\!\trans} \Big) \mathcal{W}^{\mu\alpha}_{},
\end{eqnarray}
where $\varepsilon^{\mu\nu\alpha\beta}$ is the Levi-Civita tensor.
In Appendix~B the necessary algebra including commutator and
anticommutator relations is shortly surveyed.

The tensor structure of Eqs.~(\ref{II:Eq:Scalar})~--~(\ref{II:Eq:Tensor})
becomes more clear if we split these equations into longitudinal and
transverse components with respect to the hyperplane $\sigma^{}_{n,\tau}$.
The vector and axial-vector functions are decomposed according to
\begin{equation}
\label{II:VecDecomp} \mathcal{W}_{}^{\mu} = n_{}^{\mu}\mathcal{W}_{\longi}^{}
+ \mathcal{W}_{\!\trans}^{\mu} , \qquad \mathcal{W}_{(A)}^{\mu} =
n_{}^{\mu}\mathcal{W}_{\longi (A)}^{} + \mathcal{W}_{\trans (A)}^{\mu} ,
\end{equation}
where
\begin{equation}
\label{II:VecDecomp1}
\begin{array}{l}
\mathcal{W}_{\longi}^{} = n_{\alpha}^{}\mathcal{W}_{}^{\alpha} , \qquad
\mathcal{W}_{\longi(A)}^{} = n_{\alpha}^{}\mathcal{W}_{(A)}^{\alpha},
\\[6pt]
\mathcal{W}_{\!\trans}^{\mu} = \Delta_{\ \alpha}^{\mu}\mathcal{W}_{}^{\alpha}
,\qquad \mathcal{W}_{\trans (A)}^{\mu} = \Delta_{\
\alpha}^{\mu}\mathcal{W}_{(A)}^{\alpha}.
\end{array}
\end{equation}
The tensor function can be written as
\begin{eqnarray}
\label{II:TensDecomp} & & \mathcal{W}_{}^{\mu\nu} = \left(
\mathcal{U}_{}^{\mu} n_{}^{\nu} -\mathcal{U}_{}^{\nu} n_{}^{\mu} \right) +
\mathcal{W}_{\!\trans}^{\mu\nu} ,
\nonumber\\[6pt]
& & \mathcal{U}_{}^{\,\mu} = n_{\alpha} \mathcal{W}_{}^{\mu\alpha} ,\qquad
\mathcal{W}_{\!\trans}^{\mu\nu} = \Delta_{\ \alpha}^{\mu}\Delta_{\
\beta}^{\nu} \mathcal{W}_{\!\trans}^{\alpha\beta} .
\end{eqnarray}
Then we arrive at the following set of equations:
\begin{eqnarray}
\label{II:Eq1:Scalar} & & \hspace*{-5pt} {\sf D}^{}_{\tau} \mathcal{W}^{}_{}=
-4 {\sf P}^{}_{\alpha} \mathcal{U}^{\alpha},
\\[4pt]
\label{II:Eq1:VecLong} & & \hspace*{-5pt} {\sf D}^{}_{\tau}
\mathcal{W}^{}_{\longi} = - \,{\sf D}^{}_{\!\trans\alpha}
\mathcal{W}^{\alpha}_{\!\trans} ,
\\[4pt]
\label{II:Eq1:VecTrans} & & \hspace*{-5pt} {\sf D}^{}_{\tau}
\mathcal{W}^{\mu}_{\!\trans} = {\sf D}^{\mu}_{\!\trans}
\mathcal{W}^{}_{\longi} - 2\varepsilon^{\mu}_{\ \alpha\beta\lambda}\,
n^{\alpha}\, {\sf P}^{\beta}_{} \mathcal{W}^{\lambda}_{\!\trans (A)^{}} - 4m
\mathcal{U}^{\mu},
\\[4pt]
\label{II:Eq1:PScalar} & & \hspace*{-5pt} {\sf D}^{}_{\tau}
\mathcal{W}^{}_{(P)}= 2im \mathcal{W}^{}_{\longi (A)^{}} +
2i\varepsilon^{}_{\alpha\beta\lambda\varrho}\, n^{\alpha} {\sf P}^{\beta}_{}
\mathcal{W}^{\lambda\varrho}_{\!\trans},
\\[4pt]
\label{II:Eq1:PVecLong} & & \hspace*{-5pt} {\sf D}^{}_{\tau}
\mathcal{W}^{}_{\longi(A)} = 2im \mathcal{W}^{}_{(P)^{}} - \,{\sf
D}^{}_{\!\trans\alpha} \mathcal{W}^{\alpha}_{\!\trans (A)} ,
\\[4pt]
\label{II:Eq1:PVecTrans} & & \hspace*{-5pt} {\sf D}^{}_{\tau}
\mathcal{W}^{\mu}_{\!\trans(A)^{}} = - 2 \varepsilon^{\mu}_{\
\alpha\beta\lambda}\, n^{\alpha} {\sf P}^{\beta}_{}
\mathcal{W}^{\lambda}_{\!\trans} + {\sf D}^{\mu}_{\!\trans}
\mathcal{W}^{}_{\longi (A)^{}} ,
\\[4pt]
\label{II:Eq1:TensLong} & & \hspace*{-5pt} {\sf D}^{}_{\tau}
\mathcal{U}^{\mu}= - {\sf P}^{\mu}_{}\mathcal{W} + m
\mathcal{W}^{\mu}_{\!\trans} - {\sf D}^{}_{\!\trans\alpha}
\mathcal{W}^{\mu\alpha}_{\!\trans} ,
\\[4pt]
\label{II:Eq1:TensTrans} & & \hspace*{-5pt} {\sf D}^{}_{\tau}
\mathcal{W}^{\mu\nu}_{\!\trans}= i\varepsilon^{\mu\nu\alpha\beta}_{}\,
n^{}_{\alpha} {\sf P}^{}_{\beta} \mathcal{W}^{}_{(P)} - {\sf
D}^{\mu}_{\!\trans} \mathcal{U}^{\nu} + {\sf D}^{\nu}_{\!\trans}
\mathcal{U}^{\mu} .
\end{eqnarray}
This representation for the mean-field kinetic equation will prove to be
particularly convenient in the quasi-classical limit.

\subsection{The charge conservation}
In order to describe the picture consistently the
polarization current~(\ref{II:PolCurr}), which enters the Maxwell
equations~(\ref{II:Maxwell:Tot}), must be expressed in terms of the
Wigner function.
As shown in paper I [see~(I.5.9)],
the polarization current can be written in the form
\begin{equation}
\label{II:PolCurr1}
j^{\mu}(x)=e \int \frac{d^4p}{(2\pi)^3}\,\delta(p\cdot n)\,
{\mathcal W}^{\mu}(x^{}_{\trans},p^{}_{\trans};\tau=x\cdot n),
\end{equation}
where ${\mathcal W}^{\mu}$ is the vector component~(\ref{II:Wig:Vector})
of the Wigner function. Let us prove that the fundamental charge conservation
law $\partial^{}_{\mu} j^{\mu}=0$ is satisfied in our theory.

It is convenient to define for any function
$G(x^{}_{\trans},p^{}_{\trans};\tau)$ the transformation
\begin{equation}
\label{II:Transf:G}
\overline{G}(x)=\int \frac{d^4p}{(2\pi)^3}\,\delta(p\cdot n)\,
G(x^{}_{\trans},p^{}_{\trans};\tau=x\cdot n).
\end{equation}
Then Eq.~(\ref{II:PolCurr1}) takes a compact form
\begin{equation}
\label{II:PolCurr2}
j^{\mu}(x)=e\,\overline{\mathcal W}^{\mu}(x).
\end{equation}
It follows easily from Eq.~(\ref{II:Transf:G}) that
\begin{equation}
\label{II:PropTrnsf1}
\partial^{}_{\mu} \overline{G}(x)= n^{}_{\mu}
\overline{\displaystyle\left(\frac{\partial G}{\partial \tau}\right)}
+ \nabla^{}_{\mu} \overline{G}(x).
\end{equation}
Note also that for functions $G$ which go to zero as
$|p^{}_{\trans}|\to \infty$ we have
\begin{equation}
\label{II:PropTrnsf234}
\overline{{\sf D}^{}_{\tau} G}= n^{\mu} \partial^{}_{\mu} \overline{G},
\qquad
\overline{{\sf D}^{\mu}_{\trans} G}= \nabla^{\mu}\overline{G},
\qquad
\overline{{\sf P}^{\mu} G}= \overline{p^{\mu}_{\trans}G}.
\end{equation}
Applying the transformation~(\ref{II:Transf:G}) to Eq.~(\ref{II:Eq:Vector}),
we obtain
\begin{eqnarray}
\label{II:CurrL}
n^{\lambda} \partial^{}_{\lambda} j^{\mu}=
&-&\left(n^{\mu} \nabla^{}_{\nu} - n^{}_{\nu} \nabla^{\mu} \right) j^{\nu}
\nonumber\\[8pt]
{}&-& 2e\,\varepsilon^{\mu}_{\ \alpha\beta\lambda}\, n^{\alpha}\,
\overline{\left(p^{\beta}_{\trans} {\mathcal W}^{\lambda}_{(A)}\right)}
-4em \overline{\mathcal W}^{\,\mu\nu} n^{}_{\nu}.
\end{eqnarray}
On the other hand, we may write
$$
\partial^{}_{\mu} j^{\mu} =
n^{}_{\mu} n^{\lambda}\partial^{}_{\lambda} j^{\mu} +
\nabla^{}_{\mu} j^{\mu}.
$$
Combining this with Eq.~(\ref{II:CurrL}), we see that
$\partial^{}_{\mu} j^{\mu}=0$.

One can follow a similar procedure to derive other conservation laws and
balance equations for local quantities like the mass current, the spin density,
the magnetic moment density, and the angular momentum density.
The advantage of the hyperplane formalism over the previous
approaches to the mean-field QED kinetic theory~\cite{BGR91,ShinRafelski}
is that all the balance equations and conservation laws will have a manifestly
covariant form.

\setcounter{equation}{0}

\section{The quasi-classical limit}

\subsection{The local-field approximation}
In view of practical applications of the theory, it is of interest to study
kinetic processes in QED plasmas depending on a slowly varying external
EM field. To consider this case, we insert the constants $\hbar$ and
$c$ into Eq.~(\ref{II:Wigner:Eq}) and the
operators~(\ref{II:DTau})~--~(\ref{II:PTrans}).
Then we obtain the kinetic equation
\begin{equation}
\label{II:Wigner:Eq:D}
{\sf D}^{}_{\tau}W=-{imc\over\hbar}\,\big[\gamma^{}_{\longi},W\big]
-{i\over2}\,{\sf D}^{}_{\trans\mu}
\left[S^{\mu},W\right] - {1\over\hbar}\,{\sf P}^{}_{\mu}
\left\{S^{\mu},W\right\}
\end{equation}
and the following expressions for the operators with the
  corresponding gradient expansions:
\begin{eqnarray}
\label{II:DTau:D}
{\sf D}^{}_{\tau}&=& \frac{\partial}{\partial\tau}
-{e\over c}\int\limits^{1/2}_{-1/2} ds\,
n^{\mu} {\mathcal F}^{}_{\mu\nu}
\left(\tau,x^{}_{\trans} -is\hbar\nabla^{}_{p}\right)
\nabla^{\nu}_{p}
\nonumber\\[6pt]
{}&= &
\frac{\partial}{\partial\tau} -\frac{e}{c}\,
n^{\mu}{\mathcal F}^{}_{\mu\nu}\,\nabla^{\nu}_{p}
+\frac{e\hbar^2}{24 c}
\left(\nabla\cdot\nabla^{}_{p}\right)^2
n^{\mu}{\mathcal F}^{}_{\mu\nu}\,\nabla^{\nu}_{p} +\ldots,
\\[12pt]
\label{II:DTrans:D}
{\sf D}^{}_{\trans\mu}&=&
\nabla^{}_{\mu}
- {e\over c}\int\limits_{-1/2}^{1/2} ds\,
{\mathcal F}^{}_{\trans\mu\nu}
\left(\tau,x^{}_{\trans} -is\hbar\nabla^{}_{p}\right)
\nabla^{\nu}_{p}
\nonumber\\[6pt]
{}&= &
\nabla^{}_{\mu} -\frac{e}{c}\,{\mathcal F}^{}_{\trans\mu\nu}\nabla^{\nu}_{p}
+ \frac{e\hbar^2}{24 c}
\left(\nabla\cdot\nabla^{}_{p}\right)^2
n^{\mu}{\mathcal F}^{}_{\trans\mu\nu}\,\nabla^{\nu}_{p}+\ldots,
\\[12pt]
\label{II:PTrans:D}
{\sf P}^{}_{\mu}&=&
p^{}_{\trans\mu}
-{ie\hbar\over c} \int\limits_{-1/2}^{1/2} s\,ds\,
{\mathcal F}^{}_{\trans\mu\nu}
\left(\tau,x^{}_{\trans} -is\hbar\nabla^{}_{p}\right)
\nabla^{\nu}_{p}
\nonumber\\
{}&=&
p^{}_{\trans\mu} -
\frac{e\hbar^2}{12 c}\left(\nabla\cdot\nabla^{}_{p}\right)
{\mathcal F}^{}_{\trans\mu\nu}\nabla^{\nu}_{p} +\ldots
\end{eqnarray}
The condition that the terms containing the field derivatives be small reads
\begin{equation}
\label{II:QClass:1}
\bar{\lambda}^{}_{B}\ll l^{}_{EM},
\end{equation}
where $\bar{\lambda}^{}_{B}$ is the average de Broglie wave length for
fermions and $l^{}_{EM}$ is the characteristic length for variations of the
EM field in the system. For laser induced plasmas
the latter quantity is roughly equal to the
wave length of the external laser field.
We will refer to the condition~(\ref{II:QClass:1}) as the local
approximation, from which the
operators~(\ref{II:DTau:D})~--~(\ref{II:PTrans:D})
can be concluded to be
\begin{eqnarray}
\label{II:Dtau:Class}
& &
{\sf D}^{}_{\tau}=
\frac{\partial}{\partial\tau} -\frac{e}{c}\,
n^{\mu}{\mathcal F}^{}_{\mu\nu}\,\nabla^{\nu}_{p}\, ,
\\[6pt]
& &
\label{II:Dtr:Class}
{\sf D}^{}_{\trans\mu}=
\nabla^{}_{\mu} -
\frac{e}{c}\,{\mathcal F}^{}_{\trans\mu\nu}\nabla^{\nu}_{p}\, ,
\\[6pt]
\label{II:P:Class}
& &
{\sf P}^{}_{\mu}=p^{}_{\trans\mu}\, .
\end{eqnarray}

In the local-field approximation,
Eqs.~(\ref{II:Eq1:Scalar})~--~(\ref{II:Eq1:TensTrans})
(with inserted $\hbar$ and $c$) become
\begin{eqnarray}
\label{II:EqD:Scalar}
& & \hspace*{-5pt} {\sf D}^{}_{\tau} \mathcal{W}^{}_{}=
-{4\over\hbar}\, p^{}_{\trans\mu} \mathcal{U}^{\mu},
\\[4pt]
\label{II:EqD:VecLong}
& & \hspace*{-5pt} {\sf D}^{}_{\tau}
\mathcal{W}^{}_{\longi} = - \,{\sf D}^{}_{\!\trans\mu}
\mathcal{W}^{\mu}_{\!\trans} ,
\\[4pt]
\label{II:EqD:VecTrans}
& & \hspace*{-5pt} {\sf D}^{}_{\tau}
\mathcal{W}^{\mu}_{\!\trans} = {\sf D}^{\mu}_{\!\trans}
\mathcal{W}^{}_{\longi} -
{2\over\hbar}\,\varepsilon^{\mu}_{\ \alpha\beta\lambda}\,
n^{\alpha}p^{\beta}_{\trans} \mathcal{W}^{\lambda}_{\!\trans (A)^{}}
- {4mc\over\hbar}\,
\mathcal{U}^{\mu},
\\[4pt]
\label{II:EqD:PScalar}
& & \hspace*{-5pt} {\sf D}^{}_{\tau}
\mathcal{W}^{}_{(P)}= i{2mc\over\hbar}\, \mathcal{W}^{}_{\longi (A)^{}} +
i{2\over\hbar}\,\varepsilon^{}_{\alpha\beta\lambda\varrho}\,
n^{\alpha} p^{\beta}_{\trans}
\mathcal{W}^{\lambda\varrho}_{\!\trans},
\\[4pt]
\label{II:EqD:PVecLong}
& & \hspace*{-5pt} {\sf D}^{}_{\tau}
\mathcal{W}^{}_{\longi(A)} = i{2mc\over\hbar}\,
\mathcal{W}^{}_{(P)^{}} - \,{\sf D}^{}_{\!\trans\mu}
\mathcal{W}^{\mu}_{\!\trans (A)} ,
\\[4pt]
\label{II:EqD:PVecTrans}
& & \hspace*{-5pt} {\sf D}^{}_{\tau}
\mathcal{W}^{\mu}_{\!\trans(A)^{}} =
- {2\over\hbar}\,
 \varepsilon^{\mu}_{\ \alpha\beta\lambda}\, n^{\alpha} p^{\beta}_{\trans}
\mathcal{W}^{\lambda}_{\!\trans} + {\sf D}^{\mu}_{\!\trans}
\mathcal{W}^{}_{\longi (A)^{}} ,
\\[4pt]
\label{II:EqD:TensLong}
& & \hspace*{-5pt} {\sf D}^{}_{\tau}
\mathcal{U}^{\mu}= -{1\over\hbar}\, p^{\mu}_{\trans}\mathcal{W} +
{mc\over\hbar}\,
\mathcal{W}^{\mu}_{\!\trans} - {\sf D}^{}_{\!\trans\alpha}
\mathcal{W}^{\mu\alpha}_{\!\trans} ,
\\[4pt]
\label{II:EqD:TensTrans}
& & \hspace*{-5pt} {\sf D}^{}_{\tau}
\mathcal{W}^{\mu\nu}_{\!\trans}= {i\over\hbar}\,
\varepsilon^{\mu\nu\alpha\beta}_{}\,
n^{}_{\alpha} p^{}_{\trans\beta} \mathcal{W}^{}_{(P)} -
{\sf D}^{\mu}_{\!\trans} \mathcal{U}^{\nu} + {\sf D}^{\nu}_{\!\trans}
\mathcal{U}^{\mu} .
\end{eqnarray}

It should be emphasized that the local approximation in the operators
${\sf D}^{}_{\tau}$, ${\sf D}^{\mu}_{\trans}$, and
${\sf P}^{\mu}$ does not necessarily implies that all quantum effects
are neglected. It can easily be seen from
Eqs.~(\ref{II:EqD:Scalar})~--~(\ref{II:EqD:TensTrans}) that some
components of the Wigner function show non-analytic behavior in the limit
$\hbar\to 0$ describing quantum phenomena like pair production in
strong fields~\cite{Roberts00_1,Bloch99_1}.
This aspect of the ``one-time'' mean-field theory of QED was discussed by
Bialynicki-Birula et al.~\cite{BGR91} in their study of the Dirac vacuum in
strong external fields.
For QED plasmas the situation is somewhat similar to the problem of the
Dirac vacuum, but, generally speaking, a self-consistent description of
quantum effect in plasmas involves the photon kinetics.
This point will be detailed in a special article.

\subsection{Quasi-classical kinetic equations for fermions}
We now want to obtain the quasi-classical limit of
Eqs.~(\ref{II:EqD:Scalar})~--~(\ref{II:EqD:TensTrans}) for
those components of the Wigner function which determine the
polarization current~(\ref{II:PolCurr1}).

We first notice that Eq.~(\ref{II:EqD:VecTrans}) allows to eliminate
${\mathcal U}^{\mu}$ in the other equations. In particular, we have
\begin{equation}
\label{II:pU}
p^{}_{\trans\mu} {\mathcal U}^{\mu}=
\frac{\hbar}{4mc} \,p^{}_{\trans\mu}
\left(
{\sf D}^{\mu}_{\trans} {\mathcal W}^{}_{\longi}
- {\sf D}^{}_{\tau} {\mathcal W}^{\mu}_{\trans}
\right),
\end{equation}
which is to be inserted into Eq.~(\ref{II:EqD:Scalar}).
Then we can observe that, in the quasi-classical limit
($\hbar\rightarrow 0$),
Eq.~(\ref{II:EqD:TensLong}) leads to the relation
\begin{equation}
\label{II:WTrans:Cl}
{\mathcal W}^{\mu}_{\trans}=
\frac{p^{\mu}_{\trans}}{mc}\,{\mathcal W}.
\end{equation}
Thus Eqs.~(\ref{II:EqD:Scalar}) and~(\ref{II:EqD:VecLong})
give a
closed set of quasi-classical equations for
${\mathcal W}$ and ${\mathcal W}^{}_{\longi}$:
\begin{equation}
\label{II:WW:Class}
\begin{array}{l}
\displaystyle
{\sf D}^{}_{\tau} {\mathcal W} -
\frac{1}{m^2c^2}\,p^{}_{\trans\mu}\,
{\sf D}^{}_{\tau} \left(p^{\mu}_{\trans} {\mathcal W}\right)
+ \frac{1}{mc}\,p^{\mu}_{\trans}
\,{\sf D}^{}_{\trans\mu} {\mathcal W}^{}_{\longi}=0,
\\[10pt]
\displaystyle
{\sf D}^{}_{\tau} {\mathcal W}^{}_{\longi} +
\frac{1}{mc}\,{\sf D}^{}_{\trans\mu}
\left( p^{\mu}_{\trans} {\mathcal W}\right)=0.
\end{array}
\end{equation}
With  (\ref{II:Dtau:Class}) and (\ref{II:Dtr:Class}),
it is easy to verify that in the quasi-classical limit
\begin{equation}
\label{D:Ident}
p^{}_{\trans\mu} {\sf D}^{}_{\tau} p^{\mu}_{\trans}=
-\frac{e}{c}\,n^{\mu} {\mathcal F}^{}_{\mu\nu} p^{\nu}_{\trans},
\qquad
{\sf D}^{}_{\trans\mu} p^{\mu}_{\trans}=0.
\end{equation}
Using these relations, a little algebra shows that Eqs.~(\ref{II:WW:Class})
can be written in a more symmetric form
\begin{equation}
\label{II:WW:Class2}
\begin{array}{l}
\displaystyle
{\sf D}^{}_{\tau}
\left(\frac{\epsilon(p^{}_{\trans})}{mc^2}\, {\mathcal W} \right) +
\frac{v^{\mu}_{\trans}}{c}\,
{\sf D}^{}_{\trans\mu} {\mathcal W}^{}_{\longi}=0,
\\[10pt]
\displaystyle
{\sf D}^{}_{\tau} {\mathcal W}^{}_{\longi} +
\frac{v^{\mu}_{\trans}}{c}\,
{\sf D}^{}_{\trans\mu}\left(
\frac{\epsilon(p^{}_{\trans})}{mc^2}\, {\mathcal W}
\right)=0,
\end{array}
\end{equation}
where we have introduced the dispersion relation for fermions on the
hyperplane
\begin{equation}
\label{II:Disper:Plane}
\epsilon(p^{}_{\trans})=c\sqrt{m^2c^2 -p^{2}_{\trans}}
\end{equation}
and the transverse four-velocity
\begin{equation}
\label{II:Veloc:Tr}
v^{\mu}_{\trans}= \frac{c^2}{\epsilon(p^{}_{\trans})}\,p^{\mu}_{\trans}\, .
\end{equation}
We now define the distribution functions for electrons ($w$) and
positrons ($\bar{w}$) on the hyperplane $\sigma^{}_{n,\tau}$:
\begin{equation}
\label{II:DistrSig}
\begin{array}{l}
\displaystyle
w(x^{}_{\trans},p^{}_{\trans};\tau)=
\frac{1}{2}\left\{
\frac{\epsilon(p^{}_{\trans})}{mc^2}\,
{\mathcal W}(x^{}_{\trans},p^{}_{\trans};\tau)+
{\mathcal W}^{}_{\longi}(x^{}_{\trans},p^{}_{\trans};\tau)
\right\},
\\[14pt]
\displaystyle
\bar{w}(x^{}_{\trans},p^{}_{\trans};\tau)=
\frac{1}{2}\left\{
\frac{\epsilon(p^{}_{\trans})}{mc^2}\,
{\mathcal W}(x^{}_{\trans},-p^{}_{\trans};\tau)-
{\mathcal W}^{}_{\longi}(x^{}_{\trans},-p^{}_{\trans};\tau)
\right\}.
\end{array}
\end{equation}
These functions satisfy independent kinetic equations which follow from
Eqs.~(\ref{II:WW:Class2}). In the expanded form, we have
\begin{equation}
\label{II:KE:Plane}
 \begin{array}{l}
\displaystyle
\left(\frac{\partial}{\partial\tau} + \frac{v^{\mu}_{\trans}}{c}\,
\nabla^{}_{\mu}\right)w
-\frac{e}{c}\left(
n^{\mu}{\mathcal F}^{}_{\mu\nu} +
\frac{v^{\mu}_{\trans}}{c}\,{\mathcal F}^{}_{\trans\mu\nu}\right)
\nabla^{\nu}_{p}w=0,
\\[14pt]
 \displaystyle
\left(\frac{\partial}{\partial\tau} + \frac{v^{\mu}_{\trans}}{c}\,
\nabla^{}_{\mu}\right)\bar{w}
+\frac{e}{c}\left(
n^{\mu}{\mathcal F}^{}_{\mu\nu} +
\frac{v^{\mu}_{\trans}}{c}\,{\mathcal F}^{}_{\trans\mu\nu}\right)
\nabla^{\nu}_{p} \bar{w}=0.
 \end{array}
\end{equation}
These equations are in fact nothing more than a generalization of
relativistic Vlasov equations to the case that the distribution functions
for particles and antiparticles are defined on arbitrary hyperplanes
$\sigma^{}_{n,\tau}$. In the special ``instant frame'', where
$n^{\mu}=(1,0,0,0)$ and $x^{\mu}=(ct,\vek{r})$, Eqs.~(\ref{II:KE:Plane})
take the well-known form [see Appendix~A]
 \begin{equation}
 \label{II:KE:Lab}
 \begin{array}{l}
 \displaystyle
 \frac{\partial w}{\partial t} + \vek{v}\cdot \vek{\nabla}w
 + e\left[\vek{\mathcal E} +{}{1\over c}(\vek{v}\times\vek{\mathcal B})
 \right] \cdot
 \frac{\partial w}{\partial\vek{p}}=0\, ,
 \\[14pt]
\displaystyle
 \frac{\partial \bar{w}}{\partial t} + \vek{v}\cdot \vek{\nabla}\bar{w}
 - e\left[\vek{\mathcal E} +{}{1\over c}(\vek{v}\times\vek{\mathcal B})
 \right] \cdot
 \frac{\partial\bar{w}}{\partial\vek{p}}=0\, ,
 \end{array}
 \end{equation}
where $\vek{v}=c^2\vek{p}/\epsilon(\vek{p})$
is the velocity vector, and
$\epsilon(\vek{p})=c\sqrt{\displaystyle\vek{p}^2 + m^2c^2}$ is
the relativistic energy in the ``instant frame''.

The kinetic equations~(\ref{II:KE:Plane}) describe the evolution of the
distribution functions with respect to the time-like variable $\tau$.
Note, however, that $w$ and $\bar{w}$ can also be regarded as functions of
the space-time point $x$ and the four-vector $p^{}_{\trans}$, according to
 \begin{equation}
 \label{II:wwX}
 \hspace*{-30pt}
 w(x,p^{}_{\trans})
 \equiv w(x^{}_{\trans},p^{}_{\trans};\tau=x\cdot n),
 \quad
 \bar{w}(x,p^{}_{\trans})
 \equiv\bar{w}(x^{}_{\trans},p^{}_{\trans};\tau=x\cdot n).
 \end{equation}
This interpretation of the distribution functions allows to
put Eqs.~(\ref{II:KE:Plane}) into a more compact form.
First we note that
$$
v^{\mu}_{\trans} {\mathcal F}^{}_{\trans\mu\nu}\nabla^{\nu}_{p}=
v^{\mu}_{\trans} {\mathcal F}^{}_{\mu\nu}\nabla^{\nu}_{p},
$$
which follows directly from the definition of ${\mathcal F}^{}_{\trans\mu\nu}$,
Eq.~(\ref{II:TransFTens}).
We next consider the relations
 \begin{equation}
 \label{II:Derivs}
 \begin{array}{l}
 \hspace*{-25pt}
 \displaystyle
 u^{\mu}\partial^{}_{\mu} w(x,p^{}_{\trans}) =
  u^{\mu}\nabla^{}_{\mu}w(x,p^{}_{\trans}) +
\left.
 (u\cdot n)
 \left(
 \frac{\partial}{\partial\tau} w(x^{}_{\trans},p^{}_{\trans};\tau)
 \right)\right|_{\tau=x\cdot n}\, ,
 \\[14pt]
 \hspace*{-25pt}
\displaystyle
 u^{\mu}\partial^{}_{\mu} \bar{w}(x,p^{}_{\trans}) =
  u^{\mu}\nabla^{}_{\mu}\bar{w}(x,p^{}_{\trans}) +
\left.
 (u\cdot n)
 \left(
 \frac{\partial}{\partial\tau} \bar{w}(x^{}_{\trans},p^{}_{\trans};\tau)
 \right)\right|_{\tau=x\cdot n},
 \end{array}
 \end{equation}
where the time-like unit vector $u^{\mu}$ is defined as
 \begin{equation}
 \label{II:u}
 u^{\mu}= \frac{\epsilon(p^{}_{\trans})}{mc^2}
 \left(n^{\mu} + \frac{v^{\mu}_{\trans}}{c}\right).
 \end{equation}
Elimination of the $\tau$-derivatives between Eqs.~(\ref{II:KE:Plane})
and~(\ref{II:Derivs}) and some rearrangement leads to the equations
 \begin{equation}
 \label{II:Ke:XP}
 \begin{array}{l}
 \displaystyle
 u^{\mu} \left(\partial^{}_{\mu} -{e\over c}\,{\mathcal F}^{}_{\mu\nu}(x)\,
  \nabla^{\nu}_{p}\right) w(x,p^{}_{\trans})=0\, ,
  \\[14pt]
  \displaystyle
u^{\mu} \left(\partial^{}_{\mu} +{e\over c}\,{\mathcal F}^{}_{\mu\nu}(x)\,
  \nabla^{\nu}_{p}\right) \bar{w}(x,p^{}_{\trans})=0\, .
 \end{array}
 \end{equation}
One can verify, e.g.,  by going to the ``instant frame'', that~(\ref{II:u})
is the four-velocity of a particle with the
four-momentum $p^{\mu}$ satisfying the mass-shell condition $p^2=m^2c^2$.
Equations~(\ref{II:Ke:XP}) may thus be interpreted as the evolution
equations for the distribution functions with respect to the invariant
proper time.

In order to guarantee a self-consistent description of the plasma in
the quasi-classical approximation,
the polarization current~(\ref{II:PolCurr1})
must be expressed in terms of the distribution functions $w$ and $\bar{w}$.
To do this, we recall the quasi-classical result~(\ref{II:WTrans:Cl})
and write
\begin{equation}
\label{II:WmuC1}
 {\mathcal W}^{\mu}= n^{\mu}{\mathcal W}^{}_{\longi} +
\frac{p^{\mu}_{\trans}}{mc}\,{\mathcal W}.
\end{equation}
Elimination of ${\mathcal W}^{}_{\longi}$ and ${\mathcal W}$
with the aid of Eqs.~(\ref{II:DistrSig}) gives
\begin{equation}
\label{II:WmuC2}
{\mathcal W}^{\mu}(x,p^{}_{\trans})= \frac{mc^2}{\epsilon(p^{}_{\trans})}
\left\{
u^{\mu}(p^{}_{\trans})\,w(x,p^{}_{\trans}) -
u^{\mu}(-p^{}_{\trans})\,\bar{w}(x,-p^{}_{\trans})
\right\},
\end{equation}
so that the polarization current~(\ref{II:PolCurr1})
takes the form (with the inserted Planck's constant)
 \begin{equation}
 \label{II:PolCurr3}
 j^{\mu}(x)= e \int \frac{d^4 p}{(2\pi\hbar)^3}\,
 \delta(p\cdot n)\, \frac{mc^2}{\epsilon(p^{}_{\trans})}\,
 u^{\mu}(p^{}_{\trans})
 \Big[
 w(x,p^{}_{\trans}) - \bar{w}(x,p^{}_{\trans})
 \Big].
 \end{equation}
By using Eqs.~(\ref{II:Ke:XP}), it can easily be verified that the above
expression for the current is consistent with the conservation law
$\partial^{}_{\mu} j^{\mu}=0$.

\subsection{The invariant quasi-classical distribution function for fermions}
It is interesting that the polarization current~(\ref{II:PolCurr3})
can be rewritten in a form where the four-vector $n$ does not appear.
First we notice that the delta function $\delta(p\cdot n)$ in
Eq.~(\ref{II:PolCurr3}) may be replaced by
$\delta\left(p\cdot n - \epsilon(p^{}_{\trans})/c \right)$
because other functions in the integrand do not depend on
$p^{}_{\longi}=p\cdot n$. Then, according to the identity
 \begin{equation}
 \label{II:IntIdent}
 \int d^4p\, \frac{\delta(p\cdot n - \epsilon(p^{}_{\trans})/c)}
 {\epsilon(p^{}_{\trans})}\, \left(\cdots\right)=
 \frac{2}{c} \int\limits_{p^0>0} d^4p\, \delta(p^2 -m^2c^2)
 \left(\cdots\right),
 \end{equation}
we can rewrite Eq.~(\ref{II:PolCurr3}) as
 \begin{equation}
 \label{II:PolCurr4}
 \hspace*{-25pt}
 j^{\mu}(x)= 2emc \int\limits_{p^0>0} \frac{d^4p}{(2\pi\hbar)^3} \,
 \delta(p^2 -m^2c^2) u^{\mu}(p^{}_{\trans})
 \Big[
 w(x,p^{}_{\trans}) - \bar{w}(x,p^{}_{\trans})
 \Big].
 \end{equation}
Finally, from Eqs.~(\ref{II:Veloc:Tr}) and~(\ref{II:u}) follows
 \begin{equation}
 \label{II:u1}
 u^{\mu}(p^{}_{\trans})= \frac{p^{\mu}}{mc} -
 \frac{n^{\mu}}{mc} \big[ p\cdot n - \epsilon(p^{}_{\trans})/c\big].
 \end{equation}
With the mass-shell constraint $p^2=m^2c^2$ we have
$u^{\mu}=p^{\mu}/mc$, so that Eq.~(\ref{II:PolCurr4}) becomes
 \begin{equation}
 \label{II:PolCurrInv}
 j^{\mu}(x)=2e \int \frac{d^4p}{(2\pi\hbar)^3}\, p^{\mu}
 \Big[ f(x,p) -\bar{f}(x,p)\Big]
 \end{equation}
after introducing the mass-shell distribution functions for particles
and antiparticles:
\begin{equation}
 \label{II:InvDistr}
 \begin{array}{l}
 f(x,p)= \Theta(p^{0})\,\delta(p^2 -m^2c^2)\, w(x,p^{}_{\trans}),
 \\[10pt]
 \bar{f}(x,p)=\Theta(p^{0})\, \delta(p^2 -m^2c^2)\, \bar{w}(x,p^{}_{\trans}),
 \end{array}
\end{equation}
where $\Theta(p^{0})$ is the unit step function.
The mean-field kinetic equations for these
 functions can be derived from
Eqs.~(\ref{II:Ke:XP}). We will give only the derivation of the equation for
$f(x,p)$ since the equation for $\bar{f}(x,p)$ is obtained analogously.

Multiplying the first of Eqs.~(\ref{II:Ke:XP}) by
$\Theta(p^{0})\,\delta(p^2-m^2c^2)$ and
again using the fact that on the mass-shell  $u^{\mu}=p^{\mu}/mc$, we have
 \begin{equation}
 \label{II:KEMassSh}
 \Theta(p^{0})\,\delta(p^2-m^2c^2)\,p^{\mu}\left(
 \partial^{}_{\mu} -
 \frac{e}{c}\,{\mathcal F}^{}_{\mu\nu} \nabla^{\nu}_{p}
 \right)w=0.
 \end{equation}
The transverse gradient in the momentum space, $\nabla^{\nu}_{p}$, can be
represented as
 \begin{equation}
 \label{II:TrMom:Dec}
 \nabla^{\nu}_{p} =
 \partial^{\nu}_{p} - n^{\nu}\,\frac{\partial}{\partial p^{}_{\longi}},
 \end{equation}
where $\partial^{\nu}_{p}=g^{\nu\lambda}\,\partial/\partial p^{\lambda}$ and
$p^{}_{\longi}=p\cdot n$. Since $w$ does not depend on $p^{}_{\longi}$,
the operator $\nabla^{\nu}_{p}$ in Eq.~(\ref{II:KEMassSh}) may be replaced by
$\partial^{\nu}_{p}$. Finally, using the relations
 $$
 \partial^{\nu}_{p} \left[ \Theta(p^{0})\,
 \delta(p^2 -m^2c^2)\right]= 2\Theta(p^{0})\,p^{\nu}\,
 \frac{\partial\delta(p^2 -m^2c^2)}{\partial p^2}
 $$
and $p^{\mu}{\mathcal F}^{}_{\mu\nu}p^{\nu}=0$, Eq.~(\ref{II:KEMassSh})
takes the form
\begin{equation}
\label{II:KEInvPart}
 p^{\mu} \left( \partial^{}_{\mu} -
 \frac{e}{c}\,{\mathcal F}^{}_{\mu\nu}(x)\,\partial^{\nu}_{p}
 \right)f(x,p)=0.
\end{equation}
The analogous covariant kinetic equation for antiparticles reads
\begin{equation}
\label{II:KEInvAnti}
 p^{\mu} \left( \partial^{}_{\mu} +
 \frac{e}{c}\,{\mathcal F}^{}_{\mu\nu}(x)\,\partial^{\nu}_{p}
 \right)\bar{f}(x,p)=0.
\end{equation}
Formally, Eq.~(\ref{II:KEInvPart}) coincides with the well-known relativistic
kinetic equation for charged particles in a prescribed electromagnetic field
(see, e.g.,~\cite{DeGroot80}).

Here one comment is in order. We see that the invariant distribution functions
$f(x,p)$ and $\bar{f}(x,p)$ satisfy kinetic equations~(\ref{II:KEInvPart})
and~(\ref{II:KEInvAnti}) which do not give any indication of the family of
hyperplanes $\sigma^{}_{n,\tau}$ used in the derivation of these equations.
Note, however, that a unique solution of these equations exists only if
$f(x,p)$ and $\bar{f}(x,p)$ are specified on some space-like surface $\sigma$
in Minkowski space. To formulate this ``initial condition",
we have to recall Eqs.~(\ref{II:InvDistr}) which relate the
invariant distribution functions to the functions $w$ and $\bar{w}$ defined on
the family of hyperplanes $\sigma^{}_{n,\tau}$. Lorentz invariance of
the theory manifests itself by associating $f(x,p)$ and $\bar{f}(x,p)$
with an arbitrary family of hyperplanes in order to fix the ``initial
condition''.

\section{Summary and outlook}

Based on the general density matrix approach to QED plasmas~\cite{Hoell1_01},
we have derived kinetic equations for the fermionic subsystem in the
mean-field approximation. The general mean-field expression given by
Eq.~(\ref{II:Wigner:Eq}) is a covariant generalization of the
kinetic equation derived previously by Bialynicki-Birula et al.~\cite{BGR91}.
Their result is reproduced in the ``instant frame'', where the
normal vector is given by $n^{\mu}=(1,0,0,0)$.
The covariant structure of Eq.~(\ref{II:Wigner:Eq}) is particularly convenient
for the spinor decomposition which allows to separate kinetic
processes in different channels. Another advantage of this equation is
that it can be used to perform further approximations in a covariant
form. For instance, the quasi-classical limit of the mean-field
kinetic equation is presented.

Applications of relativistic mean-field theories have been discussed in
different contexts. For instance, in heavy-ion collisions a
relativistic kinetic equation in an ``instant frame'' is solved using
the relativistic Landau-Vlasov method~\cite{Wolter1_95}. 
Present ultra-relativistic heavy-ion collisions demand a consistent
relativistic approach to nonequilibrium evolution~\cite{Roberts00_1}.
Laser-plasma interactions are most 
often treated within particle in cell (PIC)
simulations~\cite{Pukhov1_99}, which follow from classical mean-field
approximations to the relativistic kinetic equation.

Spectral information is not contained in the description presented
here. In~\cite{Heinz1_98,Heinz2_98}, for instance, 
the relation between the one-time and two-time Wigner function in the
instant frame is discussed.
Applying an energy moment expansion of the two-time Wigner function,
the one-time Wigner function is given by the lowest moment, whereas
the spectral information is contained in higher moments.

There are different ways to go beyond the approximations presented in
this paper.
Quantum corrections to the quasi-classical kinetic equation can be
taken into account by expanding
Eqs.~(\ref{II:Eq1:Scalar})~--~(\ref{II:Eq1:TensTrans}) in terms of Planck's
constant. This implies to consider non-local fluctuations in the
plasma at a length scale less than the de Broglie wavelength.
Quantum effects associated with particle-antiparticle coherence,
like for instance pair production caused by strong
fields~\cite{Roberts00_1,Bloch99_1},
can give significant corrections to the Wigner function in the
treatment of QED plasmas under extreme conditions.
Pair creation in current laser-plasma experiments is realized through
a bremsstrahlung conversion of MeV electrons into MeV photons~\cite{Shkolnikov}.
For the description of such effects the photon kinetics has to be
included self-consistently into the picture presented here.

Furthermore, improving the mean-field approximation, one can consider
collisions in the plasma.
This can be done systematically by expanding the collision terms for
the photons and electrons (compare Eq.~(I:5.34) and (I:5.35)) in terms
of the fine structure constant $\alpha$. First order effects related
to emision and absorption of photons is subject of forthcoming
studies.

\setcounter{equation}{0}
\renewcommand{\theequation}{A.\arabic{equation}}

\section*{Appendix A}

\subsection*{Kinetic equation for the matrix Wigner function
in the ``instant frame''}
We consider Eq.~(\ref{II:Wigner:Eq}) in the ``instant frame'', where
$n^{\mu}=(1,0,0,0)$ and $\tau=x^{0}=t$ ($c=\hbar=1$). Introducing
the usual space-time notation $x^{\mu}=(t,\mbox{\boldmath$r$})$, we find
$$
x^{\mu}_{\!\trans}=(0,\mbox{\boldmath$r$}),
\qquad
x^{}_{\!\trans\mu}=(0,-\mbox{\boldmath$r$}),
\qquad
p^{\mu}_{\!\trans}=(0,\mbox{\boldmath$p$}),
\qquad
p^{}_{\!\trans\mu}=(0,-\mbox{\boldmath$p$}).
$$
The transverse four-gradients with respect to space-time and
the momentum variables, $\nabla^{}_{\mu}$ and
$\nabla^{\mu}_{p}$, are written in the ``instant frame'' as
$$
\nabla^{}_{\mu}=(0,\mbox{\boldmath$\nabla$}),
\qquad
\nabla^{\mu}_{}=(0,-\mbox{\boldmath$\nabla$}),
\qquad
\nabla^{}_{\!\!p\,\mu}=(0,\mbox{\boldmath$\partial$}^{}_{p}),
\qquad
\nabla^{\mu}_{p}=(0,-\mbox{\boldmath$\partial$}^{}_{p}),
$$
where
\begin{math}
\mbox{\boldmath $\nabla$} =
\partial/\partial\mbox{\boldmath$r$}
\end{math}
and
\begin{math}
\mbox{\boldmath $\partial$}^{}_{p}=
\partial/\partial\mbox{\boldmath$p$} .
\end{math}
For definiteness, Cartesian components of all three-dimensional vectors and
gradients will be written with upper Latin indices running from 1 to 3.
For instance, $\nabla^{i}=\partial/\partial r^{i}$ and
$\partial^{i}_{p}=\partial/\partial p^{i}$.
Summation over repeated Latin indices is implied.

The total electric and magnetic fields, \mbox{\boldmath$\mathcal E$} and
\mbox{\boldmath$\mathcal B$}, are defined in the ``instant frame'' as
\begin{equation}
\label{II:EB:Def1}
\mbox{\boldmath$\mathcal E$}
= -\frac{\partial\mbox{\boldmath$\mathcal A$}}{\partial t}
-\mbox{\boldmath$\nabla$}\mbox{\boldmath$\mathcal A$}^{0},
\qquad
\mbox{\boldmath$\mathcal B$}
= \mbox{\boldmath$\nabla$}\times
\mbox{\boldmath$\mathcal A$},
\end{equation}
or,
\begin{equation}
\label{II:EB:Def2}
\mathcal{E}^{i}
= \mathcal{F}_{0i}^{},
\qquad
\mathcal{B}^{i}
= \varepsilon^{ijk} \nabla_{}^{j}\mathcal{A}_{}^{k},
\end{equation}
where $\varepsilon^{ijk}$ is the three-dimensional antisymmetric symbol with
$\varepsilon^{123}=1$. Note also that, in our notation, the non-zero
components of the tensor~(\ref{II:TransFTens}) are now given by
${\mathcal F}^{ij}_{\trans}={\mathcal F}^{}_{\trans ij}=
 - \left(\nabla^{i} {\mathcal A}^{j} -
\nabla^{j} {\mathcal A}^{i}\right)$.

Further it is easy to verify that the
operators~(\ref{II:DTau})~--~(\ref{II:PTrans}) can be written as
\begin{equation}
\label{II:NewOpers}
{\sf D}^{}_{\tau}= D^{}_{t},
\qquad
{\sf D}^{}_{\trans\mu}=(0,\mbox{\boldmath$D$}),
\qquad
{\sf P}^{}_{\trans\mu}=(0,-\mbox{\boldmath$P$}),
\end{equation}
where
\begin{eqnarray}
\label{D_t}
& &
D^{}_{t}= \frac{\partial}{\partial t}
+ e\int\limits_{-1/2}^{1/2} ds\,
\mbox{\boldmath$\mathcal E$}\left(
t,\mbox{\boldmath$r$} +is\mbox{\boldmath$\partial$}^{}_{p}
\right)\cdot\mbox{\boldmath$\partial$}^{}_{p}\, ,
\\[6pt]
& &
\label{vecD}
\mbox{\boldmath$D$}=
\mbox{\boldmath$\nabla$}
+e\int\limits_{-1/2}^{1/2} ds\,
\mbox{\boldmath$\mathcal B$}\left(
t,\mbox{\boldmath$r$}+is\mbox{\boldmath$\partial$}^{}_{p}
\right)\times\mbox{\boldmath$\partial$}^{}_{p}\, ,
\\[6pt]
& &
\label{vecP}
\mbox{\boldmath$P$}=
\mbox{\boldmath$p$}-ie \int\limits_{-1/2}^{1/2} s\,ds\,
\mbox{\boldmath$\mathcal B$}\left(
\mbox{\boldmath$r$}+is\mbox{\boldmath$\partial$}^{}_{p}
\right)\times
\mbox{\boldmath$\partial$}^{}_{p}\, .
\end{eqnarray}
Finally, in the ``instant frame'' we have
$\gamma^{}_{\longi}=\gamma^{0}\equiv \beta$, so that the
matrices $S^{\mu}$ [see Eq.~(\ref{II:SMatr})] can be written in
terms of the Dirac $\alpha$-matrices as
\begin{equation}
\label{II:SMatrNew}
S^{\mu}=(0,-i\mbox{\boldmath$\alpha$}).
\end{equation}
Putting expressions~(\ref{II:NewOpers}) and~(\ref{II:SMatrNew}) into
Eq.~(\ref{II:Wigner:Eq}), it is convenient to rewrite this equation for the
modified Wigner function~\cite{BGR91}
\begin{equation}
\label{Wigner:Modif}
\widetilde{W}=W\gamma^{0},
\end{equation}
which implies that the fermionic density operator is defined
as [cf. Eq.~(\ref{II:DensOper:Com})]
$$
\hat \rho^{}_{aa'}(\mbox{\boldmath$r$},\mbox{\boldmath$r'$})=
-{1\over2}\big[
\hat\psi^{}_{a}(\mbox{\boldmath$r$}),
\hat{\psi}^{\dagger}_{a'}(\mbox{\boldmath$r'$})
\big].
$$
Then, in terms of $\widetilde{W}$, Eq.~(\ref{II:Wigner:Eq}) becomes
\begin{equation}
\label{II:BGR:Eq}
D^{}_{t}\widetilde{W}=
-im\left[\beta,\widetilde{W}\right] -{1\over2}
\mbox{\boldmath$D$}\cdot\left\{
\mbox{\boldmath$\alpha$},\widetilde{W}
\right\}
-i\mbox{\boldmath$P$}\cdot\left[
\mbox{\boldmath$\alpha$},\widetilde{W}
\right].
\end{equation}
This is the mean-field kinetic equation derived by
Bialynicki-Birula et al.~\cite{BGR91}.

\subsection*{Quasi-classical kinetic equations in the ``instant frame''}
We now aim to show that, in the ``instant frame'', the kinetic
equations~(\ref{II:KE:Plane}) take the form~(\ref{II:KE:Lab}).
First we note that in this frame the transverse
four-velocity~(\ref{II:Veloc:Tr}) is written as
$v^{\mu}_{\trans}=(0,\vek{v})$, where $\vek{v}=c^2\vek{p}/\epsilon(\vek{p})$
is the three-dimensional velocity vector of a particle with the energy
$\epsilon(\vek{p})=c\sqrt{\displaystyle\vek{p}^2 + m^2c^2}$.
Then, using the above expressions for the transverse four-gradients in the
``instant frame'', we find that
\begin{equation}
\label{II:RelLab}
 \begin{array}{l}
 \displaystyle
 n^{\mu} {\mathcal F}^{}_{\mu\nu} \nabla^{\nu}_{p}=
 - {\mathcal F}^{}_{0i} \partial^{i}_{p} =
 - \vek{\mathcal E}\cdot \frac{\partial}{\partial \vek{p}}\, ,
 \\[6pt]
 \displaystyle
 v^{\mu}_{\trans} {\mathcal F}^{}_{\trans\mu\nu} \nabla^{\nu}_{p}=
 -v^{i} {\mathcal F}^{}_{ij} \partial^{j}_{p} =
 - \left(\vek{v}\times \vek{\mathcal B}\right)\cdot
 \frac{\partial}{\partial\vek{p}}\, .
 \end{array}
\end{equation}
Finally, in the ``instant frame" we have the obvious relation
 \begin{equation}
\label{II:DerLab}
\frac{\partial}{\partial\tau} + \frac{v^{\mu}_{\trans}}{c}\,
\nabla^{}_{\mu} = {1\over c}
\left(\frac{\partial}{\partial t} + \vek{v}\cdot \vek{\nabla}\right).
 \end{equation}
Insertion of expressions~(\ref{II:RelLab}) and~(\ref{II:DerLab})
into~(\ref{II:KE:Plane}) leads to the kinetic equations~(\ref{II:KE:Lab}).

\setcounter{equation}{0}
\section*{Appendix B}

Here we give some basic relations, which are used to transform the
matrix kinetic equation~(\ref{II:Wigner:Eq}) into
Eqs.~(\ref{II:Eq:Scalar})~--~(\ref{II:Eq:Tensor}). We follow the
notation of~\cite{Zuber}.

The totally antisymmetric Levi-Civita tensor
$\varepsilon^{\mu\nu\alpha\beta}$ is defined through even and odd permutations
of $\mu\nu\alpha\beta$ with
\begin{equation}
\label{LCivita}
\varepsilon^{0123} = -\varepsilon_{0123} = 1.
\end{equation}
The relation between co- and contravariant components follows from
the metric $g_{\mu\nu} = \mbox{diag} (1,-1,-1,-1)$

From the Dirac algebra (see e.g.~\cite{Zuber}) we can calculate the
commutator and anticommutator relations appearing in the different
spinor channels of Eq.~(\ref{II:Wigner:Eq})
\begin{eqnarray*}
& &
\hspace*{-20pt}
\left[\gamma^{}_{\mu},I\right]=0,
\qquad
\left[\gamma^{}_{\mu},\gamma^{}_{\nu}\right]=-2i\bar{\sigma}^{}_{\mu\nu},
\qquad
\left[\gamma^{}_{\mu},\gamma^{}_{5}\right]=
-2\gamma^{}_{5}\gamma^{}_{\mu},
\\[8pt]
& &
\hspace*{-20pt}
\left[\gamma^{}_{\mu},\gamma^{}_{5}\gamma^{}_{\nu}\right]
=-2g^{}_{\mu\nu} \gamma^{}_{5},
\qquad
\left[\gamma^{}_{\mu},\bar{\sigma}^{}_{\mu'\nu'}\right]
= 2i\left(
g^{}_{\mu\mu'} \gamma^{}_{\nu'} -g^{}_{\mu\nu'} \gamma^{}_{\mu'}
\right),
\\[20pt]
& &
\hspace*{-20pt}
\left[\bar{\sigma}^{}_{\mu\nu},I \right]=0,
\qquad
\left[\bar{\sigma}^{}_{\mu\nu},\gamma^{}_{5}\right]=0,
\qquad
\left[\bar{\sigma}^{}_{\mu\nu},\gamma^{}_{5}\gamma^{}_{\mu'}\right]=
2i\left(
g^{}_{\nu\mu'}\gamma^{}_{5}\gamma^{}_{\mu}
-g^{}_{\mu\mu'}\gamma^{}_{5}\gamma^{}_{\nu}
\right),
\\[8pt]
& &
\hspace*{-20pt}
\left[\bar{\sigma}^{}_{\mu\nu},\bar{\sigma}^{}_{\mu'\nu'}\right]=
-2i\left(
g^{}_{\mu\mu'}\bar{\sigma}^{}_{\nu\nu'}
+g^{}_{\nu\nu'}\bar{\sigma}^{}_{\mu\mu'}
-g^{}_{\mu\nu'}\bar{\sigma}^{}_{\nu\mu'}
-g^{}_{\nu\mu'}\bar{\sigma}^{}_{\mu\nu'}
\right),
\\[20pt]
& &
\hspace*{-20pt}
\left\{\bar{\sigma}^{}_{\mu\nu},I
\right\}=2\bar{\sigma}^{}_{\mu\nu},
\qquad
\left\{\bar{\sigma}^{}_{\mu\nu}, \gamma^{}_{\mu'}
\right\}=
2\varepsilon^{}_{\mu\nu\mu'\alpha} \gamma^{}_{5}\gamma^{\alpha},
\\[8pt]
& &
\hspace*{-20pt}
\left\{\bar{\sigma}^{}_{\mu\nu},\gamma^{}_{5}
\right\}=i\varepsilon^{}_{\mu\nu\mu'\nu'}\,\bar{\sigma}^{\mu'\nu'},
\qquad
\left\{\bar{\sigma}^{}_{\mu\nu},\gamma^{}_{5}\gamma^{}_{\mu'}
\right\}=2\varepsilon^{}_{\mu\nu\mu'\alpha}\,\gamma^{\alpha},
\\[8pt]
& &
\hspace*{-20pt}
\left\{\bar{\sigma}^{}_{\mu\nu},\bar{\sigma}^{}_{\mu'\nu'}
\right\}=
2\left(
g^{}_{\mu\mu'}g^{}_{\nu\nu'} - g^{}_{\mu\nu'}g^{}_{\nu\mu'}
\right)I
+2i\varepsilon^{}_{\mu\nu\mu'\nu'} \,\gamma^{}_{5}.
\end{eqnarray*}

\section{Acknowledgments}
The main part of this work was conducted during visits in Rostock and
Moscow. V.M.~Morozov would like to thank the ``Deutsche
Forschungsgemeinschaft'' and A.~H\"oll the ``Studienstiftung des deutschen 
Volkes'' and the ``Deutsche Forschungsgemeinschaft'' for supporting
this work.

\end{document}